\documentclass[twocolumn,prl,showpacs,amssymb,superscriptaddress]{revtex4-1}

\usepackage{graphicx}
\usepackage{amsmath}
\usepackage{amsfonts}
\usepackage{amsthm}
\usepackage{amssymb}
\usepackage{amsbsy}
\usepackage{wasysym}
\usepackage{bm}
\usepackage{mathrsfs}
\usepackage{color}
\usepackage[colorlinks=true,allcolors=blue]{hyperref}

\begin{document}
\title{Superdiffusive transport of energy in one-dimensional metals}

\author{Vir B. Bulchandani}
\affiliation{Department of Physics, University of California, Berkeley, Berkeley CA 94720, USA}
\author{Christoph Karrasch}
\affiliation{Technische Universit\"at Braunschweig, Institut f\"ur Mathematische Physik, Mendelssohnstr.~3, 38106 Braunschweig, Germany}
\author{Joel E. Moore}
\affiliation{Department of Physics, University of California, Berkeley, Berkeley CA 94720, USA}
\affiliation{Materials Sciences Division, Lawrence Berkeley National Laboratory, Berkeley CA 94720, USA}

\begin{abstract}
Metals in one spatial dimension are described at the lowest energy scales by the Luttinger liquid theory.  It is well understood that this free theory, and even interacting integrable models, can support ballistic transport of conserved quantities including energy. In contrast, realistic one-dimensional metals, even without disorder, contain integrability-breaking interactions that are expected to lead to thermalization and conventional diffusive linear response.  We argue that the expansion of energy when such a non-integrable Luttinger liquid is locally heated above its ground state shows superdiffusive behavior (i.e., spreading of energy that is intermediate between diffusion and ballistic propagation), by combining an analytical anomalous diffusion model with numerical matrix product state calculations on a specific perturbed spinless fermion chain.  Different metals will have different scaling exponents and shapes in their energy spreading, but the superdiffusive behavior is stable and should be visible in time-resolved experiments.
\end{abstract}
\maketitle

Quantum many-body systems are now, thanks to recent developments, understood to support multiple universal classes of dynamical behavior at long length and time scales.  Systems may fail to thermalize to the conventional Gibbs ensemble because there exist an infinite number of (sufficiently local) conservation laws: two well-studied examples in one spatial dimension include many-body localized systems~\cite{Serbyn13,Huse14,Vosk13,Imbrie16} and quantum integrable models~\cite{Rigol08,Barthel08}.  However, most realistic condensed matter systems do not have more than a few conservation laws, and the Gibbs ensemble or thermal state based on these is still believed to be the asymptotic state of the system.  The approach to the thermal state in such a system is usually assumed to be described either by conventional hydrodynamics, if momentum is conserved, or by diffusion.

The point of this work is to argue that a simple problem of energy transport in realistic one-dimensional metals generates a type of anomalous or nonlinear diffusion, even though the system is non-integrable, thermalizing, and described in other aspects by conventional linear response.  The Luttinger liquid is the generic metallic state of interacting one-dimensional fermions, analogous to the Fermi liquid in higher dimensions but with several fundamental differences~\cite{Giamarchi}.  The low-energy limit of the Luttinger liquid is a free bosonic theory, but real Luttinger liquids contain integrability-breaking perturbations that are responsible for thermalization.  The irrelevance of these perturbations leads to superdiffusive behavior when energy expands from an initial finite heated region into the ground state.  We study this type of rapid energy spread in part because of  experiments using laser irradiation of a small region to generate an outward flux of heat in a solid~\cite{henseldynes,photothermalmic}.  These could  be performed on spin chain materials or others where thermal transport has been argued to show signs of near-integrability, although disentangling disorder and open-system effects can be complex~\cite{chernyshyevheat,berciu}.

The problem of expansion of excitations into a region previously in the ground state has been studied in many models and received new impetus with the advent of dynamical measurements on ultracold atomic gases~\cite{Bloch}.  Two illustrative classes of possible behaviors come from considering classical physics: first, the case of free particles whose different velocities lead to dispersion, and second, the classical fluid limit in which interactions lead to nonlinear behavior and propagating wavefronts.  Both these cases lead to ballistic behavior, and some kinds of interactions in one dimension lead to integrable models that also have this ballistic property.  A third class covers diffusive behavior, for example of Brownian particles.  Diffusion implies a parametrically slower rate of spreading of either particles or energy, with finite linear-response transport coefficients.  The results presented here show that even simple, well-studied problems in quantum condensed matter physics lead to long-time scaling that is distinct from these three standard possibilities.  Note that the superdiffusion described in the present work is distinct from that known to exist in momentum-conserving many-body systems~\cite{narayan,limmer,Matveev2019,Samanta2019}, in which linear-response coefficients are not finite in the thermodynamic limit but rather diverge as a power-law in system size. 

Our presentation starts with an explicit example of a local lattice Hamiltonian that shows superdiffusive behavior and can be studied quantitatively using time-dependent density-matrix renormalization group methods.  We then present a simplified model of this behavior that is equally applicable to a broad class of one-dimensional metals, because the superdiffusive behavior originates in the continuous variation of the scaling dimensions of  irrelevant integrability-breaking operators in the Hamiltonian.  (Recall that continuous variation of the {\it electron} operator's scaling dimension leads to the well-known power-laws in electron tunneling into a Luttinger liquid~\cite{kanefisher}.)  The special aspect of energy expansion into the ground state of a realistic Luttinger liquid is that the system is never fully in the linear-response regime because of the singular zero-temperature thermal conductivity. The result is an anomalous diffusion equation with solutions of Barenblatt-Pattle type, which exhibit superdiffusive space-time scaling.
% We also discuss generalizations to incorporate charge transport and thermopower effects, and what energy expansion suggests about the range of behavior in one-dimensional quantum dynamics.

\paragraph{The microscopic model.}

For a microscopic realization of universal Luttinger liquid physics that is amenable to numerical simulation, we consider a spin-$1/2$ XXZ chain in the presence of a staggered magnetic field, with Hamiltonian $H = \sum_{i=1}^N h_i$, where
\begin{equation}
\label{model}
h_i = J S_i^x S_{i+1}^x + S_i^y S^y_{i+1}+ \Delta S_i^z S_{i+1}^z + (-1)^i h S_i^z
\end{equation}
(in the following, we set $J = a = \hbar = 1$, where $a$ denotes the lattice length scale). This model was studied in previous work~\cite{Huang2013,jaksch}, and for $\Delta \neq 0$, the staggered field can be verified to break integrability of the spin-$1/2$ XXZ chain by a level-statistics analysis. Meanwhile, the effect of the staggered field perturbation on the low-energy physics of the system can be determined via bosonization. For infinitesimal $h$, the bosonized Hamiltonian can be written as
\begin{align}
\nonumber H =& \frac{u}{2} \int_0^L dx \, \left(\Pi^2 + (\partial_x \phi)^2\right) + ch \int_0^L dx \, \cos\left(2 \sqrt{\pi K} \phi \right) \\
+ & H_\textnormal{Umklapp} + H_\textnormal{band\, curvature} + H_\textnormal{higher\,terms\,in\,$h$}
\label{effmodel}
\end{align}
where $L=Na$ and the momentum and phase degrees of freedom satisfy canonical commutation relations $[\Pi(x),\phi(y)] = i\delta(x-y)$. Here, the Luttinger parameter $K$ is given by the Bethe ansatz result $2K \cos^{-1}(-\Delta) = \pi$, and various other coupling constants can be determined exactly~\cite{Lukyanov1998}. From the scaling dimension $[h] = 2-K$, it follows that the staggered field is relevant and opens a gap for $K<2$ or $-\sqrt{2} /2 < \Delta \leq 1$. However, for $K>2$, or $-1 < \Delta < -\sqrt{2}/2$, this perturbation is irrelevant and the model remains in a gapless Luttinger liquid phase. In this paper, we focus on the latter regime. 

In what follows, it is to be understood that the effective Luttinger parameter in the gapless regime, $K(\Delta,h)$, varies continuously with $\Delta$ and $h$, and in particular will differ from the Bethe ansatz prediction for $h \neq 0$. For this reason, the values of $K$ that we quote for the specific values of $\Delta$ and $h$ considered below are obtained from an independent ground-state density-matrix renormalization group (DMRG) calculation \cite{white1992,Karrasch20122}.

\paragraph{Low-temperature hydrodynamics.}

We now consider linear response transport in the system~\eqref{model}. The DC charge and heat conductivities may be defined by the Kubo formulae~\cite{Kubo1,Kubo2,Luttinger,Kapustin}
\begin{align}
\sigma_c &= \lim_{t_M \to \infty} \lim_{N\to \infty} \frac{1}{N T} \textrm{Re} \int_0^{t_M} dt\, \langle J_c(t) J_c(0) \rangle,
\\
\sigma_h &= \lim_{t_M \to \infty} \lim_{N\to \infty} \frac{1}{N} \textrm{Re} \int_0^{t_M} dt\, \langle J_h(t) J_h(0) \rangle,
\end{align}
where the respective current operators $J = \sum_{i=1}^N j_i$ are given by the continuity equations
\begin{align}
\partial_t h_i + j_{h,i+1} - j_{h,i} = 0 &\implies J_h = i \sum_{i=2}^N [h_{i-1},h_i], \\
\partial_t S^z_i + j_{c,i+1} - j_{c,i} = 0 &\implies J_c = i \sum_{i=2}^N [h_{i-1},S^z_i].
\end{align}

A numerical study of the heat and charge conductivities $\sigma_h(T)$ and $\sigma_c(T)$ in the model \eqref{model} was performed in previous work~\cite{Huang2013}, using time-dependent, finite-temperature DMRG simulations~\cite{white1992,Schollwoeck2011, White2004,Karrasch2012,Kennes2016}. It was found that for $h \neq 0$, the AC thermal conductivity exhibits an $\mathcal{O}(h^2)$ broadening of the Drude peak arising from integrability at $h=0$. This broadening yields a finite DC thermal conductivity $\sigma_h(T)$ for $h \neq 0$. (Obtaining the value of $\sigma_h(T)$ numerically for low temperatures $T \lesssim 0.2$ appears to be beyond the present state of the art.) In the same work, it was argued that in the gapless phase of the Hamiltonian~\eqref{model}, the DC charge conductivity should scale with temperature as
\begin{equation}
\label{chargecond}
\sigma_c(T) \sim T^{\nu(K)}, \quad T \to 0,
\end{equation}
at low temperatures, with $\nu(K)$ some universal function depending only on the Luttinger parameter $K$ characterising the effective Hamiltonian \eqref{effmodel}. It was further verified that for several values of $\Delta$ and $h$ in the gapless phase, the numerical values of the Luttinger parameter $K(\Delta,h)$ and the scaling exponent $\nu(K)$ obtained from DMRG are consistent with the analytical bosonization prediction \cite{Luther74,Sirker2011}
\begin{equation}
\nu(K) = 3-2K.
\end{equation}

This is an instance of a very general scenario whereby perturbing a Luttinger liquid with an irrelevant vertex operator leads to a non-trivial power-law dependence on temperature in the low-$T$ charge conductivity, which scales continuously with the Luttinger parameter $K$. This result follows by non-perturbative resummation of the charge-current autocorrelation function, combined with a low-order Taylor expansion of the perturbation self-energy in frequency~\cite{LutherPRB74,Schulz86,OshikawaAffleck}. Although such non-perturbative resummation techniques are not directly applicable to the thermal conductivity $\sigma_h(T)$, it is natural to expect that the same phenomenon occurs, with
\begin{equation}
\label{thermalcond}
\sigma_h(T) \sim T^{\lambda(K)}, \quad T \to 0,
\end{equation}
for some exponent $\lambda(K) < 0$ that depends on $K$ and the scaling dimension of the irrelevant perturbation. For example, the assumption that $\sigma_c(T)$ and $\sigma_h(T)$ are related by Wiedemann-Franz scaling $\sigma_h(T) \sim T \sigma_c(T)$ would imply that $\lambda(K) = 1+\nu(K)$.  Indeed this holds for the tunneling electrical and thermal conductances through a single impurity in a Luttinger liquid~\cite{kanefisherthermo}, although the Lorenz number (the coefficient of the Wiedemann-Franz ratio) is modified from its Fermi liquid value.

In general, one should not assume that $\lambda(K)$ and $\nu(K)$ are always so simply related (at least it is not clear to us that this must be the case for all integrability-breaking perturbations), but even without a specific value for $\lambda(K)$, the ansatz \eqref{thermalcond} has striking consequences for the problem, mentioned in the Introduction, of expansion of a small high-temperature region into a large low-temperature background.
To see this, let us write $\sigma_h(T) = C T^{\lambda}$, where $C$ is a non-universal, temperature independent prefactor. Then in the linear-response regime and to leading order in temperature, \eqref{thermalcond} implies that temperature gradients give rise to heat currents according to $j_Q(x) \sim -C T^{\lambda} \partial_x T(x)$. We now consider states of the model \eqref{effmodel} that are in local thermodynamic equilibrium, in the sense that they are well-described by smoothly varying local temperature distribution, $T(x,t)$. For flows in such states that are driven purely by temperature gradients, the heat current coincides with the energy current, and we can write down a hydrodynamic equation 
\begin{equation}
\partial_t \rho_E= \partial_x \left(C T^{\lambda}\partial_x T\right)
\end{equation}
for small perturbations $\rho_E(x,t)$ of energy density relative to the ground state energy density, which is expected to hold to leading order in $T$ and its gradients. At low temperatures, it is also true that the temperature dependence of $\rho_E(x,t)$ is fixed by a local equation of state, of the form $\rho_E(x,t) \sim  BT(x,t)^2$, where by the low-energy properties of conformal field theories, $B=\pi k_B^2 /6 \hbar v$. This gives rise to the non-linear diffusion equation
\begin{equation}
\label{NLDE}
\partial_t \rho_E = D \partial_x^2 \left(\rho_E^m\right)
\end{equation}
for $\rho_E$, where the exponent $m$ is given in terms of $\lambda$ by $m=(\lambda+1)/2$, and the constant $D = C/2mB^m$. For a Fermi liquid with finite mean free path as $T \rightarrow 0$, $\lambda=1$ and we recover ordinary diffusion of heat. However, in the context of weakly perturbed Luttinger liquids, for which we expect that $\lambda \neq 1$ in general, more exotic scenarios can arise. If $\lambda>1$, \eqref{NLDE} is the \emph{porous medium equation}, whose solutions are characterized by subdiffusive space-time scaling, while if $\lambda<1$, this equation becomes the \emph{fast diffusion equation}, whose solutions show superdiffusive space-time scaling \cite{Vazquez2006}. Hence this model shows that the perturbed Luttinger liquid, {\it even under the assumption} that thermalization is effective enough that linear-response theory is applicable, can be expected to show superdiffusive scaling. A transparent way to see this is from the fundamental solution of \eqref{NLDE}, which for $\lambda>-1$ is the so-called ``Barenblatt-Pattle'' solution to the non-linear diffusion equation. Such solutions are characterized by a space-time scaling that varies continuously with $\lambda$,
\begin{equation}
\label{scaling}
x \sim t^\alpha, \quad \alpha = \frac{2}{\lambda+3}.
\end{equation}
Thus ``weakly perturbed'' Luttinger liquids, whose low temperature thermal conductivity exhibits the power law dependence of \eqref{thermalcond}, may exhibit a continuous range of space-time scaling exponents in their thermal transport. We now present numerical evidence for superdiffusive transport of heat, in the regime of weak integrability breaking for the Hamiltonian \eqref{model}. We find that within this model, the spreading of thermal wavepackets is characterized by a single superdiffusive exponent $2/3 < \alpha < 1$, which can be tuned by varying the strength of the integrability-breaking staggered field $h$.

While the numerical evidence shows superdiffusion, it also shows more complicated lineshapes than predicted by the locally thermalized model above, suggesting that full thermalization does not take place during the expansion. Note that collapse with a single exponent is not consistent with spreading (characterized by moments of the distribution, for example) determined by a ballistically propagating front with a weight that decays as a power-law in time, plus a central thermalized region.  Rather, there is a single limit shape that expands with a single scaling behavior.  We illustrate this behavior and then discuss its detailed relation to non-linear diffusion.

\paragraph{Numerical calculations.}

In order to demonstrate anomalous low-temperature thermal transport in Luttinger liquids, we perform DMRG simulations \cite{white1992,Schollwoeck2011} of the microscopic model \eqref{effmodel} at finite temperature \cite{White2004,Karrasch2012,Kennes2016}. The model parameters are first set to $\Delta=-0.85$ and $h=0.2$, which were found in previous work \cite{Huang2013} to generate a Luttinger liquid with effective Luttinger parameter $K \approx 2.4$. The initial data for our numerical simulation consists of a localized heated region, with inverse temperature distribution (see Materials and Methods for simulation details)

\begin{equation}
\label{initcond}
\beta(x) = \beta - (\beta-\beta_M)e^{-(x/la)^2}.
\end{equation}

\begin{figure}[t]
\includegraphics[width=0.95\linewidth]{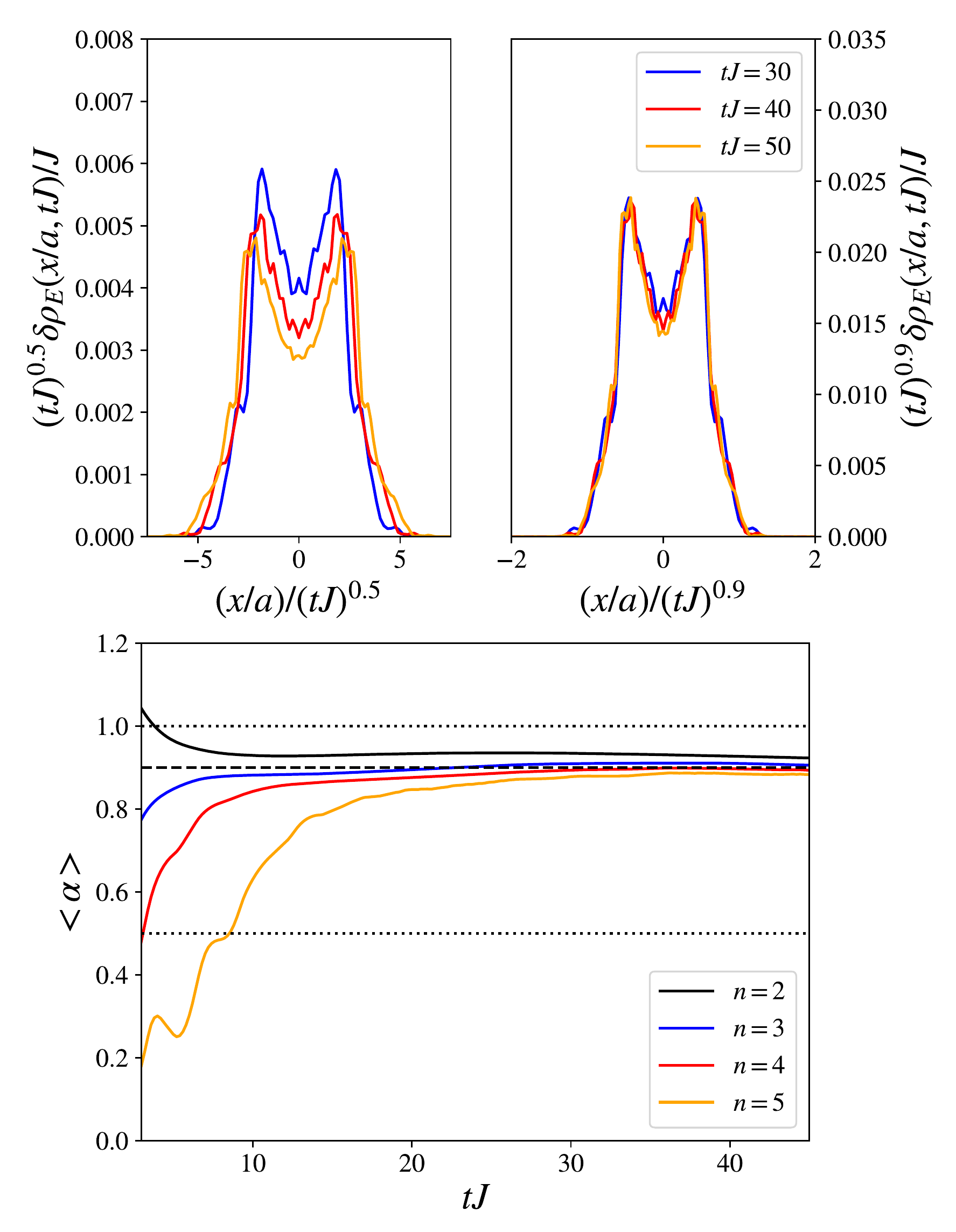}
\caption{Superdiffusion of a thermal wavepacket in a perturbed Luttinger liquid. The initial temperature profile is that of \eqref{initcond}, with $\beta J= 12$, $\beta_M J = 8$ and $l=2$. \textit{Top}: Diffusive rescaling of the wavepacket (\textit{left}) is compared with superdiffusive rescaling (\textit{right}), with exponent $\alpha \approx 0.9$. \textit{Bottom}: Logarithmic time-derivatives of the wavepacket's absolute moments indicate superdiffusion controlled by a single exponent $\alpha \approx 0.9$ (dashed line) rather than diffusive ($\alpha=0.5$) or ballistic ($\alpha=1$) scaling (dotted lines)}
\label{Fig1}
\end{figure}

In Fig.~\ref{Fig1}, we find clear evidence for superdiffusive, rather than diffusive, transport, both at the level of a naive rescaling of the thermal wavepacket, and in the scaled logarithmic time derivatives of its absolute moments, which for non-linear diffusion with a single exponent $\alpha$ should collapse to a single value at long times,
\begin{equation}
\label{logder}
\frac{1}{n} \frac {d \log{\langle |x|^n \rangle(t)}}{d \log{t}}  \to \alpha, \quad t \to \infty.
\end{equation}
Both analyses are consistent with the superdiffusive exponent $\alpha \approx 0.9$. Note that for this model, the measured value of the exponent agrees well with the Wiedemann-Franz prediction $\alpha = 2/(7-2K) \approx 0.91$.

\begin{figure}[t]
\includegraphics[width=0.95\linewidth]{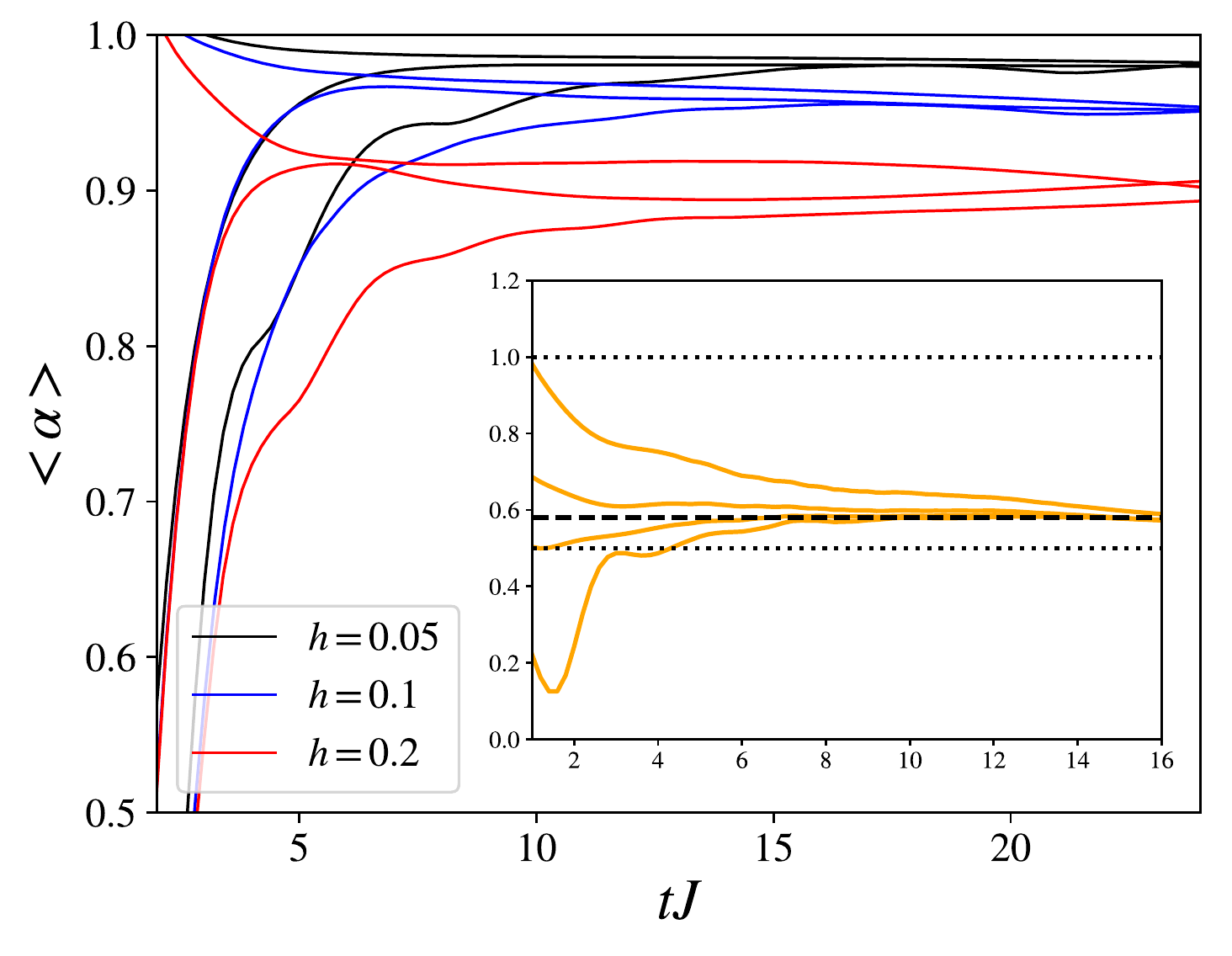}
\caption{\textit{Main figure}: Decrease of the effective superdiffusion exponent as the strength of the integrability-breaking staggered field $h$ is increased. The initial wavepacket is given by \eqref{initcond} with $\beta J = 12$, $\beta_MJ = 8$, $l=2$, the model anisotropy is set to $\Delta = -0.99$ and only $h$ is varied. Effective exponents are computed from logarithmic time derivatives of absolute moments $n=2,3,4$. \textit{Inset:} Time evolution of a higher temperature wavepacket with $\beta J=1$, $\beta_M J=0.2$, $l=2$, at anisotropy $\Delta=-0.99$ and staggered field $h=0.49$. Absolute moments $n=2,3,4,5$ demonstrate near-diffusive exponent $\alpha \approx 0.58$ (dashed line).}
\label{Fig2}
\end{figure}

We next consider the effect of varying the integrability-breaking staggered field $h$. The natural expectation is that increasing the strength of the integrability-breaking perturbation leads to a decrease in the exponent $\alpha$, bringing transport closer to normal diffusion. This is consistent with the numerical results depicted in Fig. \ref{Fig2}. For the model parameters in Fig. \ref{Fig2}, the values of the Luttinger parameter measured using DMRG~\cite{Karrasch20122} are found to be $K \sim 6-11$, which lie well outside the regime $2 < K < 2.5$ in which the Wiedemann-Franz prediction yields meaningful results. Nevertheless, the collapse to a single exponent is still consistent with the power-law assumption, \eqref{thermalcond}.

We now discuss more carefully the relation between these numerical results and the model proposed in the previous section. Strictly speaking, our model predicts that superdiffusive spreading of a localized heated region will persist indefinitely if the bulk temperature $T=0$. In the more realistic scenario of a small, non-zero bulk temperature $T>0$, we expect that wavepacket spreading will transition from superdiffusive to diffusive behaviour after some characteristic timescale $t_D(T) \sim T^{\lambda-1}$, that diverges faster than $T^{-1}$ as $T \to 0$. The temperature dependence of this timescale follows by linearizing the non-linear diffusion model \eqref{NLDE} about a constant bulk temperature. To corroborate this picture, we have checked numerically that by increasing the bulk temperature $T$, the timescale $t_D(T)$ can be brought down until the effective exponent begins to decrease towards $\alpha = 0.5$ on the numerically accessible timescale. An example for this is shown in the inset of Fig. \ref{Fig2}. (One expects that the same holds true for sufficiently shallow wave packets, but verifying this is beyond the reach of our numerics.) For the low bulk temperature $\beta J = 12$ considered in Fig. \ref{Fig1} and the main plot of Fig. \ref{Fig2}, our results indicate that the numerically accessible timescale ($t \sim 50$) is in a regime $t \ll t_D(T)$ during which the dynamics is superdiffusive. That this dynamics represents genuine anomalous diffusion, rather than a generic transient en route to diffusion, is demonstrated by the numerical observation that effective exponents obtained from different moments of the wavepacket converge to the same, superdiffusive value, as in \eqref{logder}. We have additionally checked that in the limit of bulk temperature $T=0$, for which we expect $t_D \to \infty$, superdiffusion is observed on accessible timescales. This was simulated by initializing the system in the ground state of the Hamiltonian $H'=H+\delta H$, with $H$ given by \eqref{model} and $\delta H$ a localized inhomogeneity near $x=0$, before time-evolving numerically under $H$ using pure state tDMRG~\cite{white1992,Schollwoeck2011}.

\begin{figure}[t!]
\includegraphics[width=0.95\linewidth]{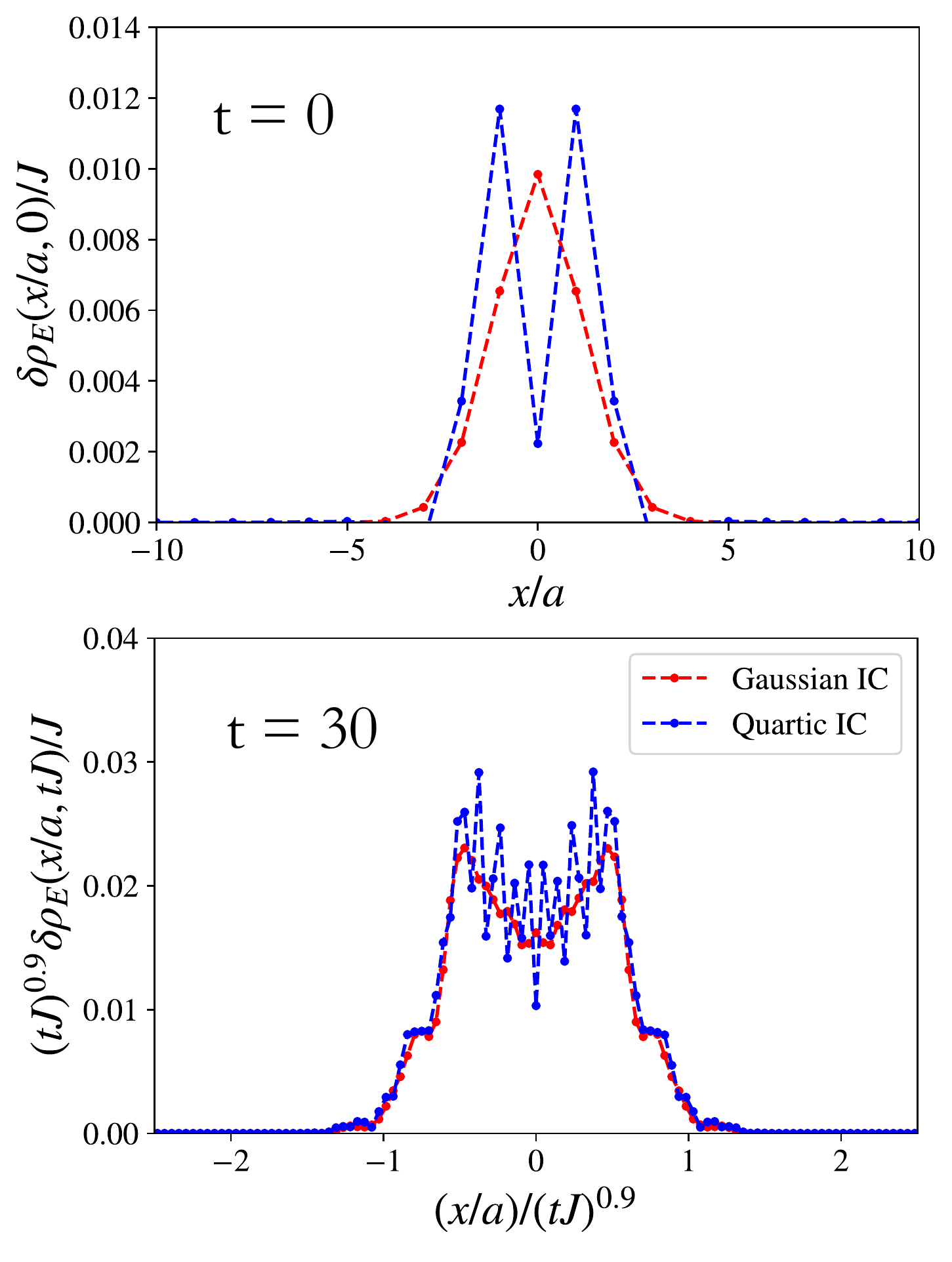}
\caption{Comparison of long-time shape of two different, localized initial profiles with the same total energy, in a model with $\Delta=-0.85$ and $h=0.2$. The Gaussian initial profile (red dash) is as in Fig. \ref{Fig1}. The quartic initial profile (blue dash) has the form $\beta(x) \propto e^{-c x^4 + d x^2}$, with $c$ and $d$ chosen to yield approximately the same total energy as the Gaussian profile. Qualitatively different initial profiles (\textit{top}) lead to a similar scaling form for the wavepackets at long times (\textit{bottom}).}
\label{Fig3}
\end{figure}

Thus the numerical collapse to a single exponent depicted in Figs. \ref{Fig1} and \ref{Fig2} indicates that our simple hydrodynamic model for propagation of heat in weakly perturbed Luttinger liquids, \eqref{NLDE}, is at least qualitatively correct, since it predicts that spreading of localized initial wavepackets at low temperature should be controlled a single superdiffusive exponent, $\alpha$. On the other hand, the splitting of the wavepacket into a doubly-peaked structure, as depicted in Fig. \ref{Fig1}, is markedly different from the shape of the Barenplatt-Pattle fundamental solution to the fast diffusion equation~\cite{Vazquez2006}, which exhibits a single maximum for all time. Moreover, the doubly-peaked structure appears to be somewhat robust to the details of the localized initial wavepacket, as shown in Fig. \ref{Fig3}. This suggests that a more refined model than \eqref{NLDE} is required to capture the precise shape of the superdiffusing wavepacket.  It is expected that using other perturbations to break integrability, or considering multi-component Luttinger liquids, will lead to different scaling functions and exponents, but the analytical model above suggests that superdiffusive or ballistic behavior should be expected as long as the scaling of linear-response thermal conductivity with temperature is larger than linear in $T$ as $T \to 0$. 

Our results are therefore consistent with a generic scenario of superdiffusive low-temperature heat transport in one-dimensional metals. A natural question is whether the same phenomenon could arise in spatial dimension $d>1$. We claim that this phenomenon \textit{can} occur in higher dimensions, but it is no longer as generic. In particular, physical systems in $d>1$ that are close to a non-interacting fixed point will exhibit superdiffusion of heat as $T \to 0$, for the same reason as the Luttinger liquid; however, in higher dimensions, such systems represent the exception when there is emergent conformal invariance rather than the rule, since the conductivity of an interacting, conformally invariant quantum critical point above one dimension is generally finite, rather than divergent as in one dimension.

If particle-hole symmetry is abandonded, the clean Fermi liquid with interactions on a lattice and a nonzero Fermi surface, which is not conformally invariant, is an example of how there can be a diverging conductivity as $T \rightarrow 0$ above one dimension.  Indeed, the conductivities diverge fast enough that an analysis of the expansion of a lump of charge and energy into the vacuum in terms of superdiffusion fails to be self-consistent. This could indicate that the ultimate behavior is ballistic, but a different method is needed for a reliable answer.

\paragraph{Discussion.} We have shown that superdiffusive spreading of heat can occur in a generic class of non-integrable, thermalizing, one-dimensional physical systems. This can be understood from a simple theoretical model, which assumes only that the temperature-dependence of the thermal conductivity in a weakly perturbed Luttinger liquid is given by a power law, $\sigma_h(T) \sim T^{\lambda}$, that diverges at low temperature.

One desirable goal for future work is a direct calculation of the low-temperature behaviour of the thermal conductivity, $\sigma_h(T)$. Analytical methods that capture the charge conductivity in a weakly perturbed Luttinger liquid~\cite{Luther74,Sirker2011} rely on the Dyson series for computing correlation functions, and do not readily generalize to the four-point functions that are required for thermal conductivities. Similarly, obtaining the power law accurately appears to be beyond the present state of the art for tDMRG methods~\cite{Huang2013}. A possible way forward is the Mori-Zwanzig memory-matrix approach; although this method is approximate in practice, it could in principle be used to estimate the thermal conductivity of a weakly perturbed Luttinger liquid~\cite{Rosch00,Jung07}.

An interesting question concerns the importance of proximate integrability in the systems under consideration. The simple anomalous diffusion model that we propose above captures the key qualitative feature of thermal wavepacket spreading in these systems, namely superdiffusion characterized by a single scaling exponent. However, the shape of the spreading wavepacket at low temperature differs from the simplest Barenblatt-Pattle form. One possible explanation for the discrepancy is that the spreading of the energy distribution and the consequent decrease of energy density violate the local thermalization assumption in the anomalous diffusion model: energy moves through a region more rapidly than the region can fully thermalize.  The recently developed hydrodynamics of quantum integrable systems~\cite{Castro-Alvaredo2016,Bertini2016} might provide a starting point for analyzing such effects, since it captures energy transport in the XXZ model without a staggered field to a remarkable degree of accuracy~\cite{Bertini2016,BVKM2017} and there is a growing understanding of how to capture integrability-breaking physics within this formalism~\cite{Friedman2019}.

Another direction for future work is to extend the current treatment to coupled charge and energy transport in systems away from half-filling, when thermopower effects become important~\cite{kanefisherthermo}. A subtlety is that the scaling of thermopower in a generic Luttinger liquid will be controlled by the leading integrability-breaking perturbation that also breaks particle-hole symmetry, which for the model \eqref{effmodel} is the band curvature correction~\cite{Pereira2007,Schulz}, distinct from the perturbation that controls thermal conductivity. At the same time, very slow relaxation of energy in this regime \cite{Matveev} suggests an intriguing possibility for ultrafast diffusion of heat.

Such refinements of the theory notwithstanding, our numerical results are consistent with an emerging understanding that for low-dimensional physical systems, the usual dichotomy between ballistic and diffusive transport can break down, in contexts ranging from classical one-dimensional systems~\cite{VanBeijeren2012,Spohn2014} to quantum integrable~\cite{Znidaric2011,Ljubotina2017,Ilievski2018,Nardis2018,Gopalakrishnan2019} as well as non-integrable~\cite{Nardis2019} models, and disordered quantum systems near the many-body localization transition~\cite{Vasseur2016}. The fact that anomalous heat transport can arise from generic perturbations to the well-studied Luttinger liquid indicates that the full richness of transport in low-dimensional quantum systems remains to be explored.

\paragraph{Materials and Methods.} Numerical results are obtained from DMRG simulations~\cite{white1992,Schollwoeck2011} of the microscopic model
\begin{equation}
\label{matmeth}
H = J \sum_{i=-N/2}^{N/2-1} S_i^x S_{i+1}^x + S_i^y S^y_{i+1}+ \Delta S_i^z S_{i+1}^z + (-1)^i h S_i^z
\end{equation}
at finite temperatures, following the method of Refs.~\cite{White2004,Karrasch2012,Kennes2016}. For the figures in the main text, we set $J = 1$, take $N=200$ sites and consider couplings $\Delta = -0.85$, $h=0.2$ in Figs. \ref{Fig1} and \ref{Fig3}, and couplings $\Delta=-0.99$, $h \in \{0.05,0.1,0.2,0.49\}$ in Fig. \ref{Fig2}. The initial state in all cases is specified by a Gaussian inverse temperature profile, 
\begin{equation}
\beta(x) = \beta - (\beta-\beta_M)e^{-(x/l)^2},
\end{equation}
parameterized by bulk ($\beta$) and central ($\beta_M$) inverse temperatures and a characteristic width $l$. The resulting profile is used to define the initial density matrix
\begin{equation}
    \rho(0) =\frac{e^{-\sum_{j=-N/2}^{N/2-1} \beta(j) h_j}}{\mathrm{tr}\left\{e^{-\sum_{j=-N/2}^{N/2-1} \beta(j)h_j}\right\}},
\end{equation}
where $h_j$ denotes the $j$-th summand in the Hamiltonian \eqref{matmeth}. This initial density matrix is then evolved in time $\rho(t) = e^{-iH t}\rho(0)e^{iHt}$, according to the time-dependent DMRG scheme proposed in Ref. ~\cite{Karrasch2012}. We use a Trotter step size of $\Delta t = 0.2/J$; the discarded weight is chosen such that the error of all quantities is at most one percent on the scale of the corresponding plot.

\textit{Acknowledgements --- } C.K. is supported by the Deutsche Forschungsgemeinschaft through the Emmy Noether program, grant no KA3360/2-2. V.B.B. acknowledges support from the DRINQS program of the Defense Advanced Research Projects Agency. J.E.M. was supported by the U.S. Department of Energy (DOE), Office of Science, Basic Energy Sciences (BES), under Contract No. DE-AC02-05-CH11231 within the Ultrafast Materials Science Program (KC2203), and a Simons Investigatorship.

% Bibliography
\bibliography{ADE}

\end{document}